\documentclass[journal=jctcce,manuscript=article]{achemso}

\usepackage[T1]{fontenc}

\usepackage{geometry}
\usepackage{setspace}
\usepackage{amssymb}

\usepackage{graphicx}
\usepackage{float}
\newfloat{scheme}{htbp}{los}
\floatname{scheme}{Scheme}
\floatname{chart}{Chart}
\newfloat{graph}{htbp}{loh}

\usepackage{chemformula} 
\usepackage[version = 4]{mhchem} 

\usepackage[numbib]{tocbibind}
\usepackage{etoolbox}
\makeatletter
\patchcmd{\acs@contact@details}{E}{*\,E}{}{}
\makeatother

\newcommand{\ZIB}{Computational Molecular Design, Zuse Institut Berlin, 14195 Berlin, Germany}
\newcommand{\FUB}{Institut für Mathematik, Freie Universit\"at Berlin, 14195 Berlin, Germany}

\title{Revealing the Atomistic Mechanism of Rare Events in Molecular Dynamics}

\author{Jakob J. Kresse}
\affiliation{\ZIB}
\email{kresse@zib.de} 

\author{Alexander Sikorski}
\affiliation{\FUB}
\alsoaffiliation{\ZIB}

\author{Marcus Weber}
\affiliation{\ZIB}
\title{Revealing the Atomistic Mechanism of Rare Events in Molecular Dynamics}

\begin{document}

\maketitle

\begin{abstract}
Interpretable reaction coordinates are essential for understanding rare conformational transitions in molecular dynamics. The Atomistic Mechanism Of Rare Events in Molecular Dynamics (AMORE-MD) framework enhances interpretability of deep-learned reaction coordinates by connecting them to atomistic mechanisms, without requiring any a priori knowledge of collective variables, pathways, or endpoints. Here, AMORE-MD employs the ISOKANN algorithm to learn a neural membership function $\chi$ representing the dominant slow process, from which transition pathways are reconstructed as minimum-energy paths aligned with the gradient of $\chi$, and atomic contributions are quantified through gradient-based sensitivity analysis. Iterative enhanced sampling further enriches transition regions and improves coverage of rare events enabling recovery of known mechanisms and chemically interpretable structural rearrangements at atomic resolution for the Müller-Brown potential, alanine dipeptide, and the elastin-derived hexapeptide VGVAPG.
\end{abstract}

\section*{Keywords}

Deep Learning; Reaction Coordinates; Explainable Artificial Intelligence; Minimum Energy Pathways; Enhanced Sampling; Koopman Operator Theory

\newpage
\section{Introduction}

Understanding the physical mechanisms that govern conformational transitions in biomolecules remains a central challenge in computational biophysics. While molecular dynamics (MD) simulations offer atomistic resolution, separating slow, collective movements from stochastic thermal fluctuations is notoriously difficult. A central question is how to identify representative transition pathways and the key atomistic motions that drive them. Traditionally, this required expert-driven identification of collective variables (CVs), such as interatomic distances, torsion angles, or root-mean-square displacements as a low dimensional description of the process \cite{bhakat2022collective}. Capturing these rare events reliably often necessitates enhanced sampling strategies of these CVs, since unbiased simulations might fail to sufficiently explore transition pathways on accessible timescales \cite{henin2022enhanced}.

Several principled approaches for discovering collective variables have been developed. 
Semi-supervised or self-supervised methods such as Robust Perron Cluster Analysis (PCCA+)~\cite{deuflhard2005robust}, 
variational approaches to conformational dynamics (VAC)~\cite{nuske2014variational}, 
and Spectral Gap Optimization of Order Parameters (SGOOP)~\cite{tiwary2016spectral} 
extract slow dynamical modes by utilizing approximative eigenfunctions of the backward generator~$\mathcal{L}$. 
Deep learning extensions of these methods, including the variational approach for Markov processes (VAMPnets)~\cite{mardt2018vampnets} and other machine-learning approaches for collective variable discovery~\cite{sultan2018automated, bonati2020data, rydzewski2021multiscale, ketkaew2022deepcv, monroe2022learning, sipka2023constructing, sun2022multitask, ray2023deep, dietrich2023machine, majumder2024machine}, have demonstrated remarkable success in identifying slow molecular modes and accelerating sampling. Nevertheless, their highly nonlinear architectures and large parameter counts often render a direct chemical interpretation challenging.
Meanwhile, the field of explainable artificial intelligence (XAI) has developed techniques such as gradient-based saliency attribution maps~\cite{ancona2019towards,adebayo2018sanity}, which make it possible to probe how neural networks arrive at their decisions. First applications of XAI in molecular kinetics have shown that post-hoc explanations can highlight which features drive a learned reaction coordinate~\cite{Kikutsuji2022ExplainingRC, naleem2023exploration, okada2024unveiling, kawashima2025investigating}. 
Complementary to such post-hoc analyses, intrinsically interpretable architectures have been developed, including models that identify compact, human-readable subsets of molecular descriptors~\cite{hooft2021discovering} and, more recently, geometric graph neural networks that learn descriptor-free collective variables with explicit node-level attribution of atomic relevance~\cite{zhang2024descriptor}.

Beyond the domain of learned representations, classical theoretical frameworks also aim to uncover mechanistic insight from molecular dynamics. Transition Path Theory (TPT)~\cite{vanden2006transition} relates rates to committor functions, but directly computing committors in high-dimensional systems is intractable. String and nudged elastic band methods~\cite{weinan2002string, weinan2005finite,henkelman2000climbing} yield clear atomistic pathways once suitable CVs and endpoints are chosen, yet their interpretability depends on these predefined coordinates. The main challenge, therefore lies in uncovering mechanistic insight without requiring any a priori specification of collective variables, endpoints, or pathways.

Here, we present the Atomistic Mechanism Of Rare Events in Molecular Dynamics (AMORE-MD) framework, designed to enhance the interpretability of deep-learned reaction coordinates at the chemical and atomic levels. In this work, we apply it in combination with the ISOKANN algorithm~\cite{rabben2020isokann}, which learns a smooth membership function~$\chi$ approximating the dominant eigenfunction of the backward operator~$\mathcal{L}$. While the Molecular Kinetics through Topology (MoKiTo)~\cite{donati2025topological} framework combines ISOKANN with topological clustering to construct global kinetic networks and identify multiple pathways in given datasets, AMORE-MD focuses on extracting local mechanistic information directly from the $\chi$-function without enforcing clustering or predefined projections, while also enhancing sampling of rare events.

We can extract mechanistic information in two complementary ways. First, by integrating along the gradient of~$\chi$ under orthogonal energy minimization, we obtain a representative trajectory, the $\chi$-minimum-energy path ($\chi$-MEP), which follows the dominant kinetic mode without requiring predefined collective variables, endpoints, a string of initial states or explicit reparameterization. Second, we analyze the gradients of~$\chi$ with respect to its inputs, providing sensitivity maps that identify which atomic distances or coordinates contribute most strongly to changes in the reaction coordinate ($\chi$-sensitivity). The $\chi$-MEP can further be used to initialize new simulations, enabling iterative sampling and retraining of~$\chi$ to improve coverage of rare transition states. This strategy is conceptually related to the recently developed path-committor-consistent artificial neural networks (PCCANN)~\cite{megias2025iterative}, which iteratively refine a committor-consistent string. In contrast to PCCANN, our framework requires neither predefined boundary sets nor an initial path guess, making it particularly suitable when no a priori mechanistic information is available.

This combination of slow-mode learning and gradient analysis allows AMORE-MD to bridge ensemble and single-path perspectives. The ensemble view arises from averaging the gradients over the stationary ensemble, which captures statistically meaningful atomic contributions across the thermodynamic landscape in the $\chi$-sensitivity, while the single-path view is represented by the $\chi$-MEP, providing a smooth and physically interpretable trajectory through conformational space. Together, these two levels of interpretation link machine-learned reaction coordinates directly to mechanistic insight. 

We validate this framework in three representative systems. First, the Müller-Brown potential demonstrates that the $\chi$-MEP recovers the zero-temperature string in a controlled benchmark. Second, alanine dipeptide in vacuum tests the method in a molecular setting with well-understood metastabilities. Finally, the hexapeptide VGVAPG in implicit solvent serves as a realistic proof of concept for larger, biologically relevant conformational transitions with multiple transition tubes. Through these examples, AMORE-MD illustrates how deep-learned reaction coordinates can be made transparent and chemically interpretable, combining statistical and mechanistic understanding within a single framework.

\subsection{Koopman operator Theory}

The dynamical behavior of a system can be described in terms of operators acting on observable functions \(f : \Omega \to \mathbb{R}\). 
In molecular dynamics, \( \Omega \subset \mathbb{R}^{3n} \) denotes the state space, where \( n \) is the number of atoms. 
The time evolution of an observable is governed by the infinitesimal generator \( \mathcal{L} \),
\[
\frac{\partial f_t(\mathbf{x})}{\partial t} = \mathcal{L} f_t(\mathbf{x}),
\]
which can be regarded as the continuous analogue of a rate matrix in discrete state space.  
The corresponding Koopman operator
\[
\mathcal{K}_{\tau} = \exp(\mathcal{L}\tau)
\]
propagates observables in time,
\[
f_{t+\tau}(\mathbf{x}) = \mathcal{K}_{\tau} f_t(\mathbf{x}) = \mathbb{E}[f_t(\mathbf{x}_{t+\tau}) \,|\, \mathbf{x}_t = \mathbf{x}],
\]
where the conditional expectation can be estimated from short burst simulations. The constant function \(\Psi_0 = 1\) is an eigenfunction of the Koopman operator with eigenvalue $1$. 
A non-trivial eigenfunction \(\Psi_1\) with the next largest eigenvalue captures the slowest relaxation process. From these a membership function \(\chi : \Omega \to [0,1]\) can be constructed as a linear combination of $\Psi_0$ and $\Psi_1$, representing the grade of membership to one of two fuzzy sets $A$ or $B$.  

The dynamics of $\chi$ is governed by the macroscopic rate equation
\begin{equation}
    \mathcal{L}  \chi = -\epsilon_1\; \chi + \epsilon_2(1-\chi)
    \label{eq:MacroRate}
\end{equation}

for the one membership function $\chi$ considered here, also derived in the supporting information.
 Thus \(\chi(\mathbf{x})\approx1\) and \(\chi(\mathbf{x})\approx0\) represent perfect membership to $A$ and $B$, while \(\chi(\mathbf{x})\approx0.5\) correspond to transition-states \(\mathbf{x}\). Therefore the membership functions represent the macro-states associated with these metastabilities and provide insights into the slowest dynamic of a system, with macroscopic rates $\epsilon$ directly available from eq (\ref{eq:MacroRate}) \cite{sikorski2024capturing}.

\subsection{ISOKANN}
ISOKANN~\cite{rabben2020isokann} extends the classical von~Mises iteration~\cite{mises1929praktische}, 
which successively applies a linear operator and normalizes the result to converge toward its dominant eigenfunction.  
A direct fixed-point iteration of the Koopman operator, however, collapses to the trivial constant function.  
To prevent this, ISOKANN replaces the normalization step by a simple affine shift--scale transformation~\(S\) 
and learns a bounded membership function \(\chi_\theta : \Omega \to [0,1]\) using a neural network (NN) with trainable parameters~\(\theta\).  

The network parameters are optimized in a self-supervised manner by iteratively minimizing the loss function
\[
\mathcal J(\theta)
=
\big\|
\chi_\theta
-
S\,\mathcal{K}_\tau \chi_{\theta-1}
\big\|^2
\]
over molecular configurations separated by a lag time~\(\tau\).  
This iterative procedure yields a stable membership function that distinguishes the dominant metastable regions of the system, serving as a low-dimensional reaction coordinate capturing the slowest process and solves eq (\ref{eq:MacroRate}).

We want to increase the interpretability of deep learned reaction coordinates, here $\chi$-functions, by extracting the most probable zero temperature transition pathways aligned with the slowest process and determining the atomic movements driving the ensemble of transition pathways. 

\section{Method}
\label{sec:Method}
We present a method for uncovering the Atomistic Mechanism Of Rare Events in Molecular Dynamics (AMORE-MD), providing mechanistic insight into slow molecular transitions. 
Our approach delivers a representative transition trajectory for the slowest dynamical process, with sensitivity resolved both along the reaction coordinate and at atomic resolution.

The method consists of four main steps:
\begin{enumerate}
    \item Run molecular dynamics simulations of the system of interest.
    \item Train a smooth neural membership function \(\chi\), which captures the slowest reaction coordinate.
    \item Approximate the minimum energy path aligned with the slowest process ($\chi$-MEP).
    \item Identify atomistic and feature saliency via the $\chi$-sensitivity measure \(\langle \|\nabla_i \chi\|^2 \rangle_z\).
\end{enumerate}

We enhance rare-event discovery through an iterative sampling scheme consisting of the first three steps: After an initial cycle of MD simulation, $\chi$-training, and $\chi$-MEP extraction, we restart MD trajectories from $\chi$-MEP states. The resulting Koopmann samples are merged with the original set to continue training $\chi$. The cycle is repeated until convergence, thereby improving coverage of transition pathways.
Although $\chi$-MEP extraction and atom-wise relevance analysis are independent, they synergize for interpretability: the $\chi$-MEP yields a compact transition trajectory, while the $\chi$-sensitivity measure \(\langle \|\nabla_i \chi\|^2 \rangle_z\) resolves atomistic contributions at finite temperature.

\subsection{A representative pathway of the slowest process}

The Minimum Energy Path (MEP) is the most likely trajectory a molecule will follow when transitioning from state $A$ to $B$  in the zero temperature limit \cite{weinan2002string}. The level-sets of the committor function $q:\mathbb{R}^{3n} \to \mathbb{R}$ lie normal to the MEP near the transition state \cite{vanden2006transition}. The committor gives the probability that a trajectory started in a state reaches the manually defined set $B$ before reaching $A$ and solves:
\[ \mathcal{L} q = 0,\]
   
with boundary conditions
\[
q(\mathbf{x}) = 
\begin{cases}
1, & \text{if } x \in B, \\
0, & \text{if } x \in A.
\end{cases}
\]
Given a sufficient spectral gap, i.e. in the case of a free energy barrier separating $A$ and $B$ much larger than the thermal energy, the committor is a monotone transform of the dominant eigenfunction $\Psi_1$ of $\mathcal{L}$ \cite{schutte2013metastability}.
As shown in the supporting information $\chi = a \Psi_0 + b \Psi_1$ with constant function $\Psi_0$ and therefore: 
\[\nabla \chi = b\nabla\psi_1\]

But this also implies that in the zero temperature regime along the transition barrier:
\begin{equation}
    \nabla\chi \propto \nabla\Psi_1\propto\nabla q
    \label{eq:monotonic}
\end{equation}
and therefore the level-sets of $\chi$ and $q$ coincide. We exploit this to define a practical procedure for approximating the MEP. Starting from an initial state $\mathbf{x}_0$ one performs energy minimization orthogonal to $\nabla\chi$ thereby recovering a state on the MEP. An Euler step along $\nabla\chi$,
\begin{equation}
    \mathbf{x}_{i+1} = \mathbf{x}_i + \epsilon\frac{\nabla\chi}{\|\nabla\chi\|},
\end{equation} 
allows to reach the next level-set on which the orthogonal energy minimization ensures staying on the MEP. This will iteratively trace the MEP until $\chi(\mathbf{x}_n)\approx 1$ is reached. The same procedure following $-\nabla\chi$ traces the path to  
$\chi(\mathbf{x}_m)\approx0$. 

We do not employ the zero-temperature $\chi$-function, but instead train on finite-temperature simulation data. In the presence of a sufficiently large spectral gap, as is typically the case in metastable MD systems, this approach can still approximate the MEP, but we will use the term $\chi$-MEP to distinguish from classical methods. In any case, the $\chi$-MEP should be interpreted as a representative of an ensemble of thermally fluctuating transition pathways along the slowest process, rather than a unique microscopic trajectory.

\subsection{Gradients of membership functions reveal atomistic sensitivity}

The membership function $\chi$ learned by ISOKANN represents the slowest dynamical process of a molecular system. 
Its gradient $\nabla \chi$ measures how infinitesimal perturbations of the atomic coordinates change the value of $\chi$, thereby indicating local kinetic sensitivity. 
For each atom~$i$, the gradient norm
\[
\left\|\nabla_i \chi\right\| = 
\left\|\left( 
\frac{\partial \chi}{\partial x_i}, 
\frac{\partial \chi}{\partial y_i}, 
\frac{\partial \chi}{\partial z_i} 
\right)\right\|
\]
quantifies the influence of atomic displacements on the learned reaction coordinate. 
Projecting $\|\nabla_i \chi\|$ onto the molecular structure yields a saliency map of atomistic relevance along the transition.

To obtain a statistically meaningful view, we average the squared gradients over the Boltzmann ensemble,
\[
\left\langle \|\nabla_i \chi\|^2 \right\rangle_\rho 
= \int_\Omega \|\nabla_i \chi(\mathbf{x})\|^2\, \rho(\mathbf{x})\, dx,
\]
with $\rho(\mathbf{x}) \propto e^{-V(\mathbf{x})/k_BT}$ the stationary distribution. 
This ensemble average captures contributions from all thermally accessible transition pathways, rather than a single trajectory, and thus reflects the collective character of the slow process. A more detailed, reaction-coordinate-resolved picture is obtained by conditioning on the level sets of~$\chi$. 
Defining the marginal density 
\[
P(z) = \int_\Omega \rho(\mathbf{x})\, \delta(\chi(\mathbf{x}) - z)\, dx,
\]
the level-set average of the atomic sensitivity becomes
\begin{equation}
    \left\langle \|\nabla_i \chi\|^2 \right\rangle_z 
= \frac{1}{P(z)} 
\int_{\Sigma_z} 
\|\nabla_i \chi(\mathbf{x})\|^2\, 
\rho(\mathbf{x})\, 
\delta(\chi(\mathbf{x}) - z)\, dx,
\end{equation}
where $\Sigma_z = \{\mathbf{x} \in \Omega \mid \chi(\mathbf{x}) = z\}$. 
Large values of $\langle \|\nabla_i \chi\|^2 \rangle_z$ indicate that displacement of atom~$i$ drives progress through the corresponding region of the reaction coordinate and thus contributes strongly to the overall transition. In addition to coordinate-based analysis, the gradient can also be evaluated directly at the model level when $\chi = \nu \circ f$, 
where $f$ denotes the molecular featurizer and $\nu$ the neural network. 
The quantity $\langle \|\nabla_i \nu\|^2 \rangle_\rho$ then provides a feature-space analogue of atom-wise saliency, identifying which internal coordinates dominate the learned slow mode.

\section{Results}
\label{sec:Results}
\subsection{Müller-Brown potential}

\begin{figure}[H]
		\centering
		\includegraphics[scale=0.063]{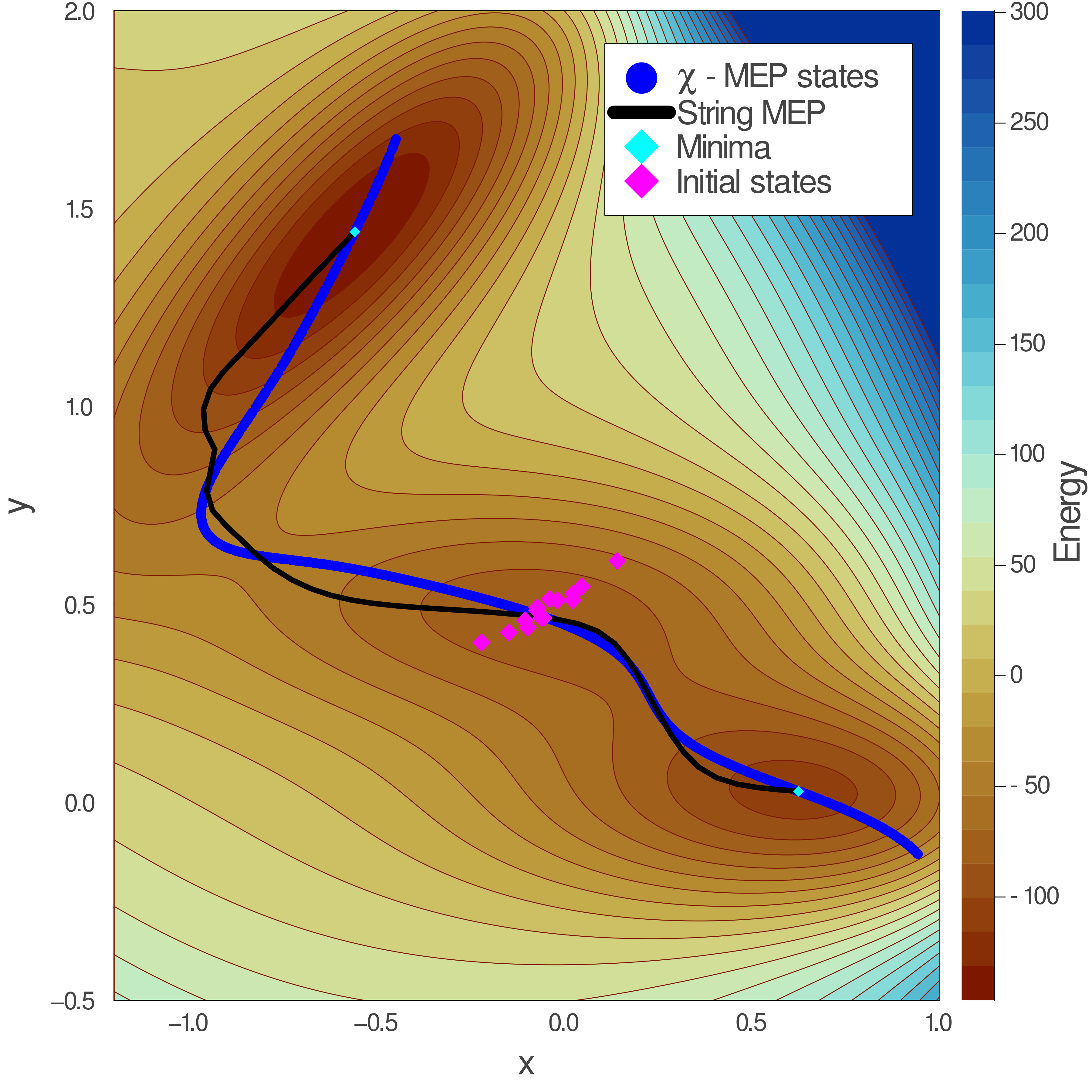} 
        \includegraphics[scale=0.063]{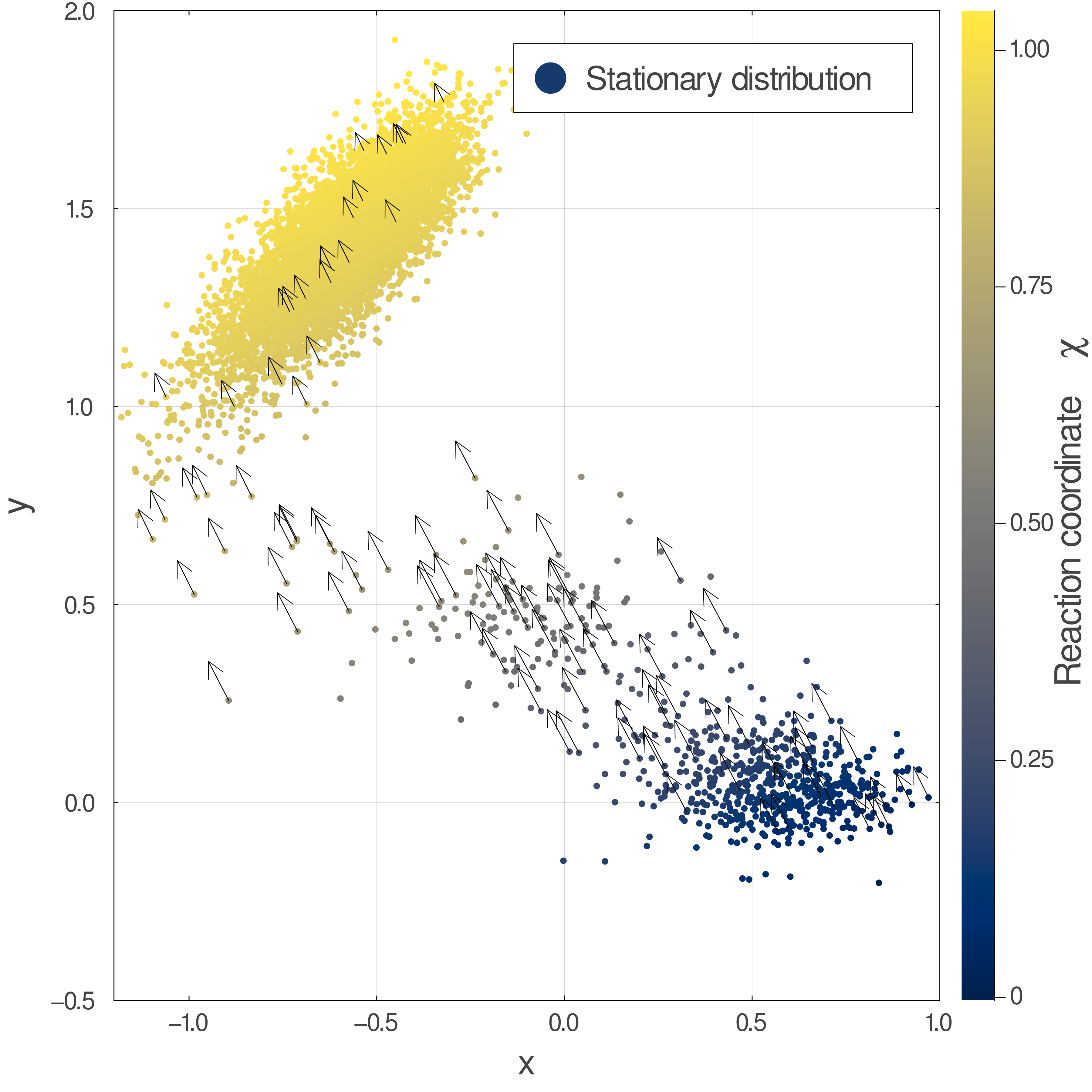}
		\caption[Transition pathways and deep learned gradients in a toy system]{\textbf{Transition pathways and deep learned gradients in a toy system.} 
        The Müller-Brown potential energy landscape is shown as a heatmap of its two coordinates (left). 
        Local minima are shown in cyan, the string MEP in black, and the $\chi$-MEP in blue, initialized from the magenta diamonds. 
        On the right, the gradient of the learned membership function $\chi$ is displayed as downscaled arrows for a subsample of the training data points (colored by their reaction coordinate value). 
        Larger gradient magnitudes highlight the regions of highest $\chi$-sensitivity along the transition barrier.}
    \label{f:mb_coord_gradient}
\end{figure}

As an illustrative benchmark, we first apply our approach to the two-dimensional Müller-Brown potential energy surface \cite{muller1979location}.  
This system exhibits two metastable minima separated by a potential barrier, making transitions between them rare events.  
The conformational landscape, together with the identified pathways, is shown in Fig.~\ref{f:mb_coord_gradient}.  

For comparison we display the result of the string method \cite{weinan2002string}, initialized as a straight line between the minima endpoints shown in cyan and discretized into 50 states. This chain of states converges to a minimum energy pathway (MEP) shown as a black curve. The $\chi$-MEP is obtained by following the gradient of the membership function $\chi$ with orthogonal energy minimization, ensuring that the path remains aligned with the learned slow coordinate.  
Both pathways traverse the minima and saddle point, and while they are overall similar, the $\chi$-MEP is slightly less skewed and cuts through somewhat higher-energy regions.  The initial states used for the $\chi$-MEP are shown as magenta diamonds, with the pathways indicated in blue, where all are following a single dominant trace. 

The learned membership function $\chi$ smoothly separates the metastable basins, taking values close to 0 in one basin and close to 1 in the other.  
Importantly, the exact values $\chi=0$ and $\chi=1$ are typically not reached inside the minima but only in the infinite-time limit.  
On the right-hand side of Fig.~\ref{f:mb_coord_gradient}, the gradients of $\chi$ are displayed as downscaled arrows for a subsample of training points.  
These arrows indicate the local direction of slow progress along the transition pathway.  
Notably, the gradient magnitudes are largest away from the metastable basins, hinting at the transition region. 
This simple example provides intuition for our approach: by combining learned membership functions, MEP-extraction and gradient-based sensitivity, we obtain both representative transition pathways and a natural measure of relevant transition steps.  

\subsection{Alanine dipeptide }
\begin{figure}[H]
		\centering
		\includegraphics[scale=0.79]{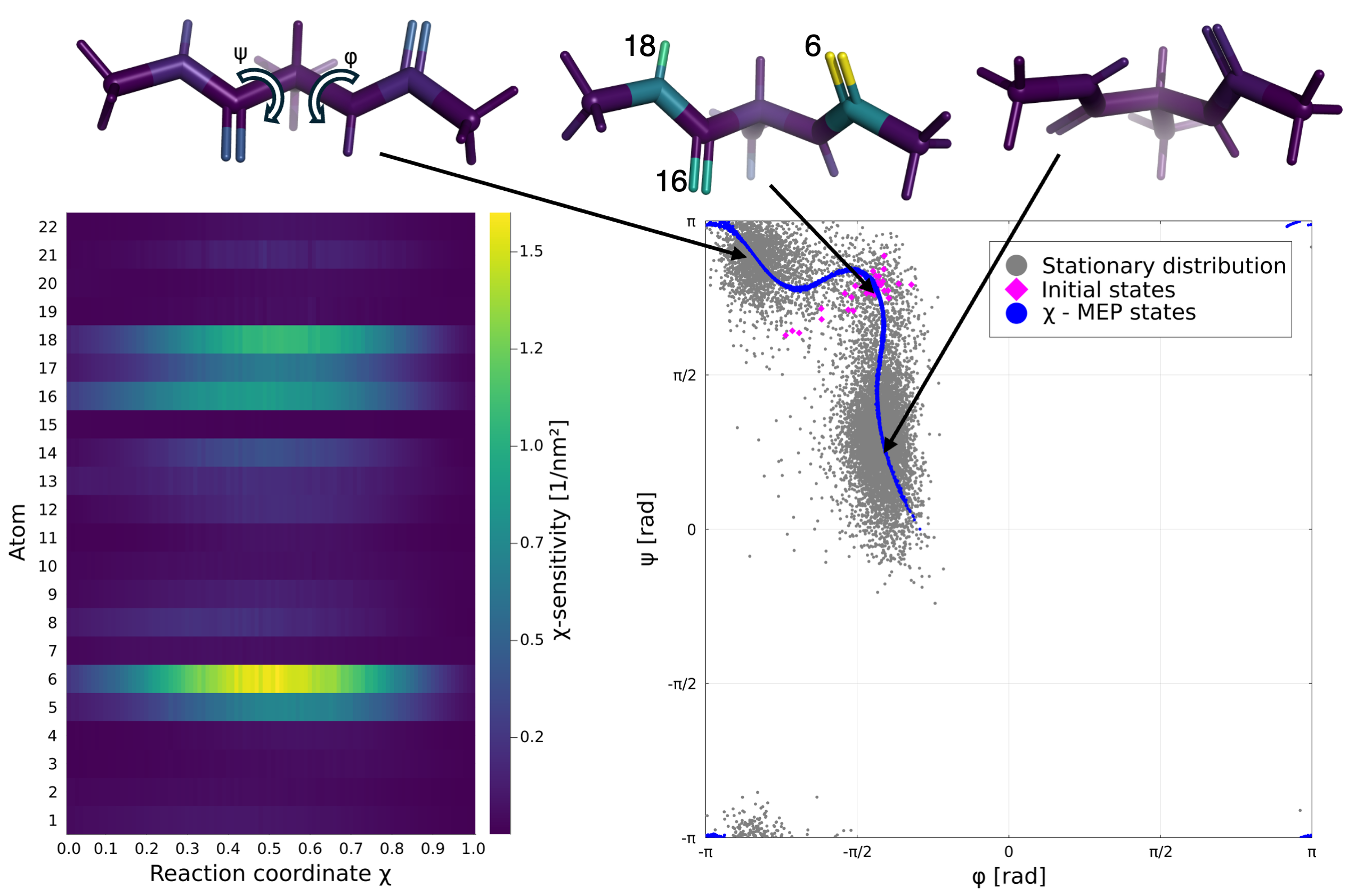} 

		\caption[Visualization of the Gradient of Alanine Dipeptide]%
        {\textbf{Conformational transitions and atomistic sensitivities in alanine dipeptide.}
        The conformational landscape of alanine dipeptide is projected onto the Ramachandran dihedral angles ($\phi$, $\psi$), revealing two dominant metastable states (bottom right). 
        The slowest kinetic mode is learned as a smooth neural membership function $\chi$. 
        The $\chi$-MEP, initialized in the magenta diamonds, is shown in blue and follows the equilibrium distribution (gray), which forms a characteristic tube-like structure around the path. 
        Three representative conformations along the $\chi$-MEP corresponding to $\chi$ values 0.1, 0.5, and 0.95 are displayed above the landscape (left to right). 
        The atomistic contributions to the learned mode are identified via the derived $\chi$-sensitivity measure. This level-set averaged squared norm of the gradient components, $\langle \|\nabla_i \chi\|^2 \rangle_z$, highlights when and where specific atomic movements contribute most to the transition. 
        These contributions are shown as a heatmap (bottom left) and further projected onto the representative structures for interpretability.}

    \label{f:diala_coord_gradient}
\end{figure}

Alanine dipeptide can undergo rare conformational transitions via rotation around its backbone dihedral angles $\phi$ and $\psi$. The most prominent transition involves a peptide bond flip that is associated with a high free energy barrier, making it a prototypical rare event in molecular dynamics simulations \cite{mironov2019systematic}. After training, ISOKANN captures the dominant slow transition coordinate $\chi$ separating the two metastable basins (Fig.~\ref{f:diala_coord_gradient}). The $\chi$-MEP traced along this learned reaction coordinate aligns with the equilibrium probability density, forming a characteristic tube in Ramachandran space. Notably, 34 initial states sampled around the transition region ($\chi \in [0.49, 0.51]$) all evolve along an indistinguishable path in $(\phi,\psi)$ space, further confirming that $\chi$ captures the dominant transition channel.

Along all transition pathways, AMORE-MD reveals the atom-wise contributions to the transition rate via the $\chi$-sensitivity. These level-set averaged squared gradient norms $\langle \|\nabla_i \chi\|^2 \rangle_z$, are computed by back-propagating through the neural membership function $\chi = \nu \circ f$, with neural network $\nu$ and featurizer $f$ mapping to pairwise distances. The resulting heatmap shows maximal contributions from backbone atoms involved in the dihedral transition, in particular atoms 6, 16, and 18, consistent with the mechanistic expectation for a peptide bond rotation. The central barrier region (around $\chi \approx 0.5$) exhibits the largest gradient magnitudes, indicating the location and character of the kinetic bottleneck as the formation of a hydrogen-bond between the amide hydrogen atom 6 and carbonyl oxygen atom 18.

\subsection{VGVAPG}
VGVAPG adopts $\beta$-turn conformations, primarily stabilized by a salt bridge between the C-terminal carboxyl group of glycine and the N-terminal amino group of valine, along with an internal hydrogen-bonding network. Computational evidence suggest that VGVAPG follows multiple distinct transition pathways, rather than a single dominant route \cite{floquet2004structural}. This makes it a representative system for studying complex biomolecular dynamics, where the ensemble of transitions cannot be captured by a single trajectory.

Our method does not aim to reconstruct a globally optimal pathway. Instead, we identify representative transitions within each tube and combine them with ensemble-weighted atom-resolved sensitivity attribution. This yields mechanistic insights into the slowest process that go beyond any one trajectory.
We identify rotation of the backbone dihedral $\psi$ of the central valine as a kinetic bottleneck for VGVAPG-folding in implicit solvent. $\psi$-values close to zero radians represent closed states of the hexapeptide, while $\psi$ values close to $\pi$ radians are open states.

\begin{figure}[H]
		\centering
		\includegraphics[scale=0.8]{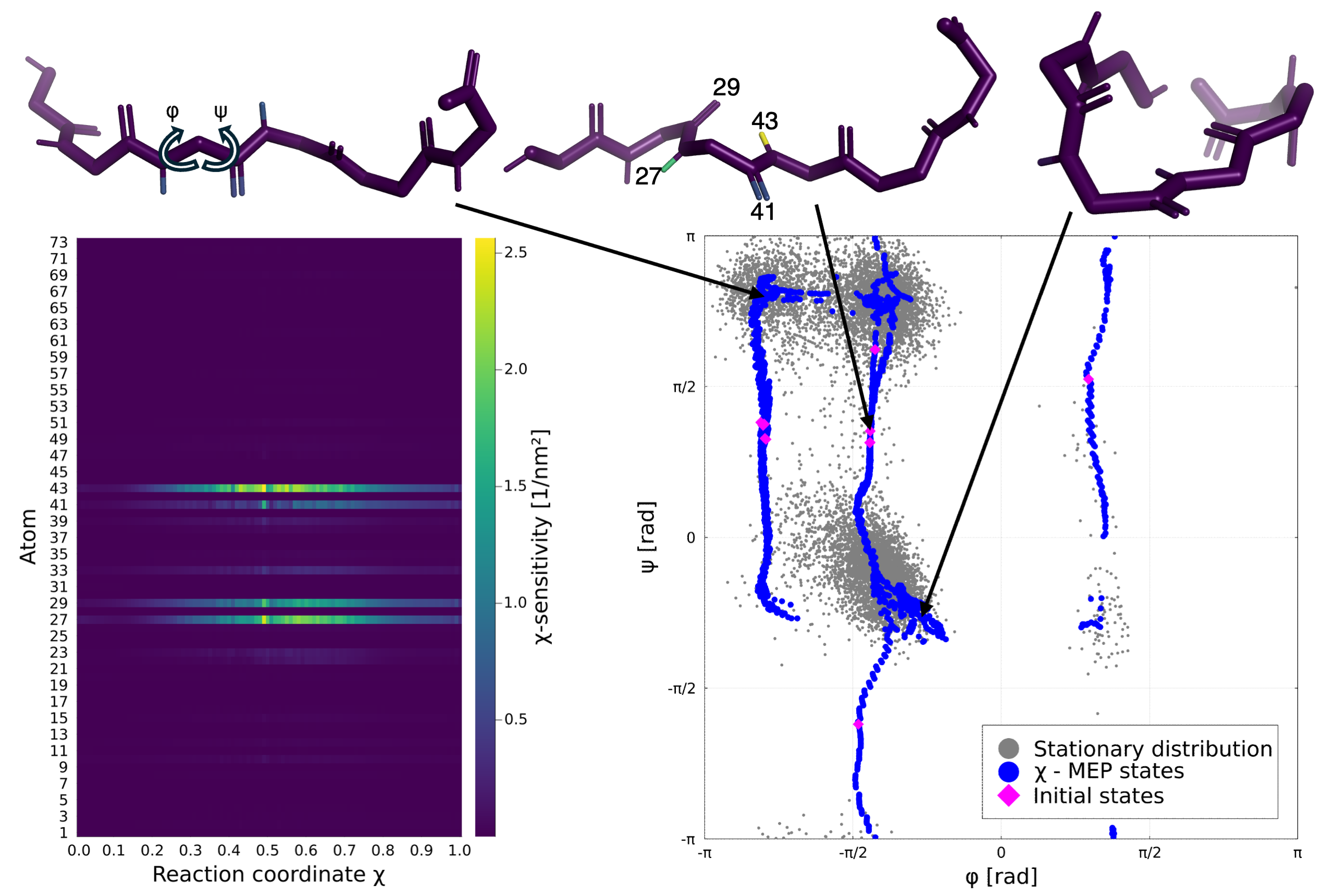} 
     \caption[Conformational transitions and atomistic sensitivities in VGVAPG]%
    {\textbf{Conformational transitions and atomistic sensitivities in VGVAPG.} 
    The conformational landscape of the elastin-mimetic peptide VGVAPG is projected onto the central valine dihedrals $\phi$ and $\psi$ (right). 
    The equilibrium distribution is shown in gray, initial states are marked in magenta, and the $\chi$-MEP states after adaptive sampling are displayed in blue. 
    Multiple transition channels are visible, reflecting the heterogeneous ensemble of conformational pathways. 
    The $\chi$-sensitivity measure $\langle \|\nabla_i \chi\|^2 \rangle_z$ reveals localized atomic regions with high influence on the learned collective mode, with strongest responses for atoms 27, 29, 41, and 43 corresponding to the $\psi$ rotation of Val2 (left).
     For clarity, representative structures above the plots are shown only for the pathway with the highest stationary density, illustrating the associated backbone rearrangements (see supplementary Movies~S2-S5 for all four pathways).
}

    \label{f:vgvapg_coord_gradient}
\end{figure}

We identify $\psi$ through the $\chi$-sensitivity $\langle \|\nabla_i \chi\|^2 \rangle_z$, which is dominated by 
the Val2 backbone atoms 27, 29, 41 and 43 (the $\psi$ dihedral itself is defined by the directly attached 
atoms 26, 28, 40 and 42). As an orthogonal degree of freedom we visualize the $\phi$ dihedral of the same 
residue and identify at least four dominant transition channels from 11 initial states 
(Fig.~\ref{f:vgvapg_coord_gradient}). \textit{Although multiple structurally and energetically distinct transition channels are accessible, the associated mechanistic abstraction is similar across them.} The Val2 backbone has to rearrange first, followed by internal hydrogen-bond formation and, lastly, salt-bridge reorganization. 
For clarity, we therefore display only the pathway with the highest stationary density around it in Fig.~\ref{f:vgvapg_coord_gradient}.

The results reported here were obtained after 100 generations of our adaptive training scheme 
(see section \ref{sec:Method}). This iterative procedure, where new Koopmann samples are initialized from $\chi$-MEP 
states and added to the training data until convergence, allows the network to extrapolate rare events more 
effectively. As a result, discontinuities in sparsely sampled regions are greatly reduced, 
and the $\chi$-MEPs shown in Fig.~\ref{f:vgvapg_coord_gradient} provide relatively continuous pathways 
across the transition channels, without explicit reparameterization. 

To assess how representative the $\chi$-MEPs are of the global mechanism, we compared the local gradients 
$\|\nabla_i \chi\|^2$ evaluated along each $\chi$-MEP to the ensemble-averaged sensitivities 
$\langle \|\nabla_i \chi\|^2 \rangle_z$. The resulting mean squared error is below 
$0.01\,\frac{1}{\mathrm{nm^2}}$for all pathways except the channel at positive $\phi$ values, 
indicating that $\chi$-MEPs generally provide a faithful atomistic representation of the 
ensemble-averaged mechanism.

Saliency attribution based on coordinates highlights the central valine backbone atoms, and chemical intuition allows one to recognize the dihedral $\psi$ as the learned slow CV. However, this identification can also be made more directly. While the featurizer $f$ is usually chosen to map to pairwise distances, it can instead map to all 10 backbone dihedral angles. Training a $\chi = \nu \circ f$-function on these features allows the central valine $\psi$ to be identified directly, by evaluating the ensemble-averaged squared gradient of the neural network $\nu$ with respect to the features, $\langle \|\nabla_i \nu\|^2 \rangle_\rho$. In this representation, the central valine $\psi$ has a value of 0.0162 $\frac{1}{rad^2}$, over one order of magnitude larger than the second-highest feature and more than two order of magnitudes larger than most other features, which lie below 0.0002 $\frac{1}{rad^2}$. 
\label{section_results}

\section{Discussion}
AMORE-MD increases the interpretability of deep-learned reaction coordinates by linking them to representative pathways and ensemble-averaged atomistic sensitivities. Across the systems considered, the framework consistently identifies physically meaningful transition mechanisms without the need to prespecify collective variables.

On the Müller-Brown surface, the $\chi$-MEP and the string MEP traverse the same metastable regions and barrier, yet deviate slightly in regions of high curvature. This difference can be attributed to the smoothness and regularization imposed during neural training, which bias $\chi$. Despite these deviations, the $\chi$-MEP remains faithful in a thermally averaged sense and captures the transition region reliably.

In alanine dipeptide at 310 K, AMORE-MD recovers the backbone-centered peptide-bond flip as the slow process analyzed in our study. A third metastable basin is known to exist on much longer time scales, but we do not observe it here. By focusing on the faster process at moderately high temperature, we deliberately probe the limits of the infinite spectral-gap assumption: while timescale separation is imperfect, the learned $\chi$ still yields a robust mechanistic description and atom-resolved sensitivity profile for the dominant transition \cite{mironov2019systematic}.

In VGVAPG, studied in implicit solvent, multiple transition channels emerge, each characterized by rearrangements of the central valine backbone. AMORE-MD consistently identifies the valine $\psi$ dihedral as the separating CV, in agreement with computational expectations \cite{floquet2004structural}. Despite the presence of several dominant transition channels, our method reveals that they all follow a common mechanistic pattern. Through this, AMORE-MD captures how diverse dynamical pathways can be represented by a shared set of defining atomistic motions. The use of implicit solvent underlines the proof-of-concept nature of this application and does not yet constitute a contribution to protein folding studies, but it demonstrates the applicability of AMORE-MD to more complex peptides.

Several caveats apply to the machine-learning component. Gradient-based sensitivities may reflect correlations rather than direct causality. For example, directly attached atoms can appear relevant because their positions are highly correlated, as seen in VGVAPG. In addition, the network may overemphasize the valine backbone due to correlated features. While recognizing such artifacts can be diagnostically valuable for improving network training, in our case retraining with dihedral features recovered the valine $\psi$ as the dominant signal, indicating that the essential mechanism was correctly captured.

AMORE-MD provides a unified framework that connects deep-learned reaction coordinates with their associated atomistic mechanisms, without requiring any a priori knowledge of the system. It identifies kinetically relevant coordinates, yields representative pathways aligned with the slowest process, and reveals sensitivities at atomic resolution. Iterative sampling further enhances the method by improving coverage of rare-event regions and stabilizing the training of the reaction coordinate through repeated cycles of simulation, learning, and resampling. While the present work focuses on proof-of-concept systems, the approach is general and scalable. Furthermore, AMORE-MD is not tied to membership functions learned by ISOKANN but can be applied to deep-learned Koopman eigenfunctions or committors in general. By linking self-supervised learning with atomistic interpretation, AMORE-MD makes it possible to uncover how slow molecular processes occur and which atomic motions drive them, providing a practical route to mechanistic understanding and design in complex chemical systems.

\newpage
\section*{Supporting Information}

Supporting Information: Theoretical motivation, methodological details, simulation parameters (PDF), and five sensitivity-colored $\chi$-MEP movies for alanine dipeptide and VGVAPG (MP4).

\section*{Acknowledgements}
This work was funded by the Deutsche Forschungsgemeinschaft (DFG)
through the Collaborative Research Center SFB~1114 “Scaling Cascades in Complex Systems”,
Projects~A05 and~B03.
We thank Dr.~Surahit Chewle, Dr.~Vikram Sunkara and Prof. Dr.~Bettina Keller for insightful discussions.


\providecommand{\latin}[1]{#1}
\makeatletter
\providecommand{\doi}
  {\begingroup\let\do\@makeother\dospecials
  \catcode`\{=1 \catcode`\}=2 \doi@aux}
\providecommand{\doi@aux}[1]{\endgroup\texttt{#1}}
\makeatother
\providecommand*\mcitethebibliography{\thebibliography}
\csname @ifundefined\endcsname{endmcitethebibliography}  {\let\endmcitethebibliography\endthebibliography}{}

\newpage

\rule{0.05in}{1.75in}%
\begin{minipage}[b][1.75in]{3.25in}
  \sffamily
  \frenchspacing

  \begin{figure}[H]
		\centering
		\includegraphics[scale=1]{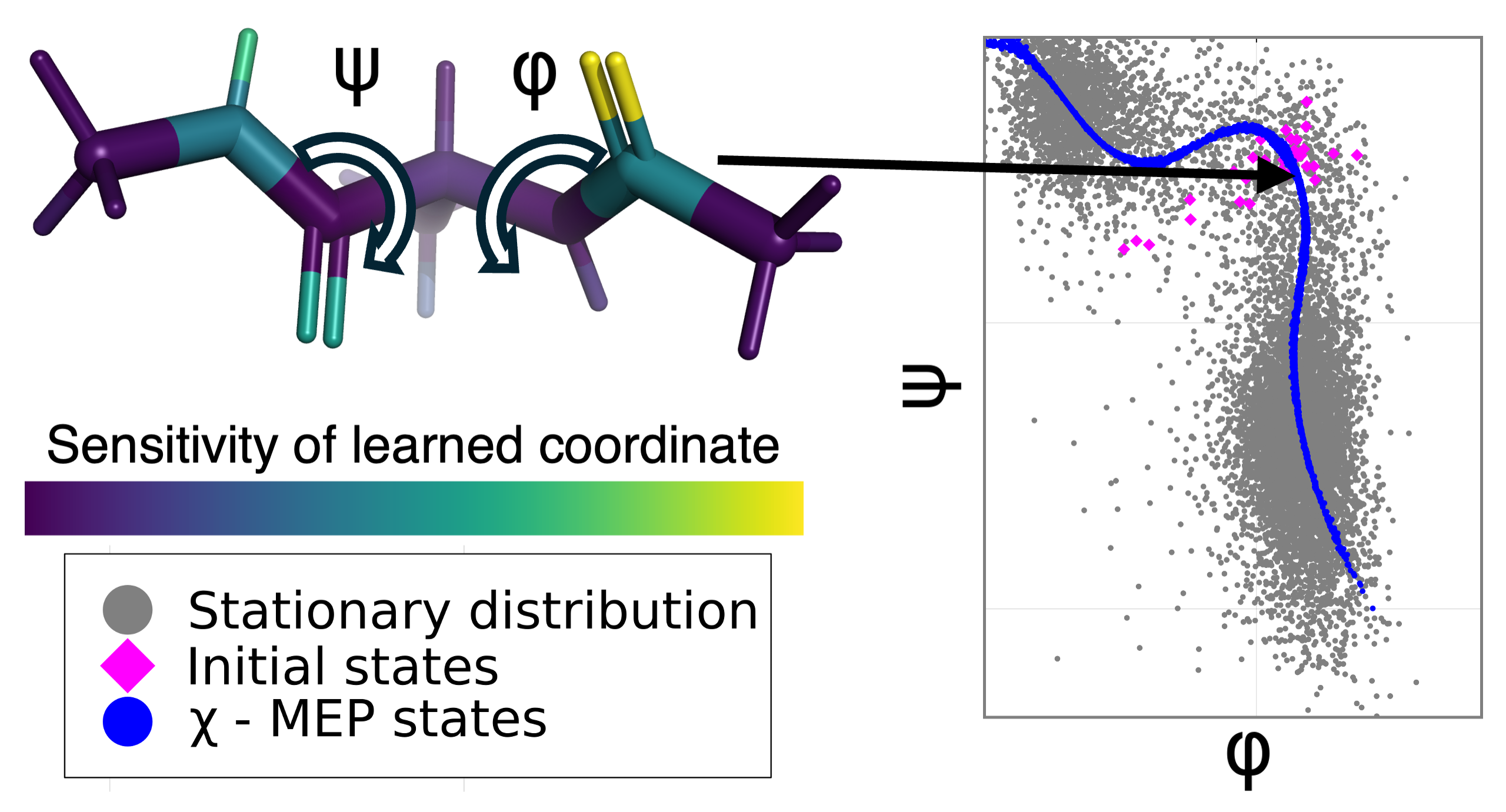} 
		\caption[For Table of Contents Only]{\textbf{For Table of Contents Only} }
    \label{f:TOC}
\end{figure}

\end{minipage}%
\rule{0.05in}{1.75in}

\end{document}